\documentclass[aps,prb,floatfix,superscriptaddress,amsmath,amssymb,noshowpacs,twocolumn]{revtex4}
\usepackage{graphics,lscape,graphicx,epsfig,amsmath,amssymb,epstopdf,wasysym,comment}
\usepackage[T1]{fontenc}
\usepackage{frcursive}
\usepackage{color}
\newenvironment{frcseries}{\fontfamily{frc}\selectfont}{}

\usepackage{subfigure}
\allowdisplaybreaks

\newcommand{\be}{\begin{equation}}
\newcommand{\ee}{\end{equation}}
\newcommand{\bea}{\begin{eqnarray}}
\newcommand{\eea}{\end{eqnarray}}
\newcommand{\ba}{\begin{eqnarray*}}
\newcommand{\ea}{\end{eqnarray*}}

\newcommand{\bR}{\mathbf{R}}

\newcommand{\bK}{\mathbf{K}}

\newcommand{\bk}{\mathbf{k}}

\newcommand{\br}{\mathbf{r}}
\newcommand{\bnot}{\mathbf{0}}

\newcommand{\dis}{\displaystyle}

\newcommand{\fract}[2]{\frac{\dis #1}{\dis #2}}

\newcommand{\eqn}[1]{(\ref{#1})}
\newcommand{\ket}[1]{\mid\! #1\rangle}
\newcommand{\bra}[1]{\langle #1\!\mid}

\newcommand{\bw}{\begin{widetext}}
\newcommand{\ew}{\end{widetext}}

\newcommand{\bGamma}{{\boldsymbol{\Gamma}}}
\newenvironment{eqs}%
{\begin{equation} \begin{aligned}}%
{\end{aligned} \end{equation} }
\newcommand{\beal}{\begin{eqs}}
\newcommand{\eal}{\end{eqs}}
\newcommand{\esp}[1]{\text{e}^{#1}}
\newcommand{\bM}{\mathbf{M}}

\begin{document}

\title{Emergent $\text{D}_6$ symmetry in fully-relaxed magic-angle twisted bilayer graphene}
\author{M.~Angeli}
\affiliation{International School for
  Advanced Studies (SISSA), Via Bonomea
  265, I-34136 Trieste, Italy} 
  \author{D.~Mandelli}
\affiliation{Department of Physical Chemistry, School of Chemistry, The Raymond and Beverly Sackler Faculty
  of Exact Sciences and The Sackler Center for Computational Molecular and Materials Science, Tel
  Aviv University, Tel Aviv 6997801, Israel}
  \author{A.~Valli}
  \affiliation{International School for
  Advanced Studies (SISSA), Via Bonomea
  265, I-34136 Trieste, Italy} 
  \affiliation{CNR-IOM Democritos, Istituto Officina dei Material, Consiglio Nazionale delle Ricerche} 
\author{A.~Amaricci}
\affiliation{International School for
  Advanced Studies (SISSA), Via Bonomea
  265, I-34136 Trieste, Italy} 
\author{M.~Capone}
\affiliation{International School for
  Advanced Studies (SISSA), Via Bonomea
  265, I-34136 Trieste, Italy} 
  \affiliation{CNR-IOM Democritos, Istituto Officina dei Material, Consiglio Nazionale delle Ricerche} 
  \author{E.~Tosatti}
\affiliation{International School for
  Advanced Studies (SISSA), Via Bonomea
  265, I-34136 Trieste, Italy} 
\affiliation{CNR-IOM Democritos, Istituto Officina dei Material, Consiglio Nazionale delle Ricerche} 
\affiliation{International Centre for Theoretical Physics (ICTP), Strada Costiera 11, I-34151 Trieste, Italy}
\author{M.~Fabrizio}
\affiliation{International School for
  Advanced Studies (SISSA), Via Bonomea
  265, I-34136 Trieste, Italy} 

\begin{abstract}
We 
present
a tight-binding calculation of a twisted bilayer graphene at magic angle $\theta\sim 1.08^\circ$,  
allowing for full, in- and out-of-plane, relaxation of the atomic positions. The resulting band structure displays as usual four narrow mini bands around the neutrality point, well separated from all other bands after the lattice relaxation.  
A thorough 
analysis of the mini-bands Bloch functions reveals 
an emergent $D_6$ symmetry, despite the 
lack of 
any manifest point group symmetry in the
relaxed lattice.
The Bloch functions 
at the $\Gamma$ point 
are degenerate in pairs, reflecting the so-called valley degeneracy. Moreover,  
each of them is 
invariant under 
C$_{3z}$, i.e., 
transforming like one-dimensional,
in-plane symmetric
 irreducible representation of 
an "emergent"
$D_6$ group. 
Out of plane,
the lower doublet is even under C$_{2x}$, while the upper doublet is odd, which implies that at least eight Wannier orbitals, two $s$-like and two $p_z$-like for each 
of the two supercell sublattices AB and BA 
are necessary, 
probably 
not sufficient, to describe the 
four
mini bands.  
This unexpected 
one-electron
complexity is likely to play an important role in the still unexplained metal-insulator-superconductor phenomenology of this system.

\end{abstract}

\date{September 28, 2018}                                          

\maketitle

\section{Introduction}
The discovery of the insulating behaviour in small angle twisted bilayer graphene (tBLG), \cite{Herrero-1,Yankowitz} and the appearance of superconducting domes upon slight hole- or 
electron-doping those insulating phases,~\cite{Herrero-2,Yankowitz}
has stimulated an intense theoretical effort to understand this phenomenon. 
At small "magic" angles $\theta \approx 1.1^{\circ}$, the electronic structure of tBLG 
is characterized by four extremely narrow bands, with a bandwidth of $\approx 10$~meV, 
which lie around the charge neutrality point in the reduced Brillouin zone 
of the emergent moir\`e superlattice.~\cite{MacDonald-PNAS2011}
Specifically, at charge neutrality these bands are half-filled, and thus one would expect 
an insulating behaviour upon adding either four holes or four electrons per moir\`e unit cell, 
as indeed observed experimentally. 
In reality, tight-binding calculations,\cite{Trambly2,Shallcross,Sboychakov} as well as 
more reliable electronic structure approaches based on DFT,\cite{Trambly,Morell,Bernevig_topo} 
show that when the graphene layers are kept rigid the mini bands 
around the magic angles are not always separated from other bands at the $\bGamma$ point, 
in contrast with experiments. 
However, once the tBLG lattice is allowed to relax,\cite{Nam_Koshino_PRB,Kaxiras} 
even the simple tight-binding scheme shows a relatively large gap opening, 
which separates the flat 
mini-bands
from 
all 
others.
Experimentally,
there is 
additional
evidence \cite{Yankowitz} of an insulating behaviour also when one or three holes/electrons are injected with respect to neutrality. 
Because of that and of 
the very 
non-dispersive
character of the mini bands, 
it is tempting to invoke an important role of strong electronic correlations.\cite{Herrero-1} 
The common approach 
dealing
with strong correlations is adding electron-electron repulsion on top 
of a tight-binding lattice model.
However, the large number of atoms contained in the unit cell 
(up to $\approx 11,000$ at $\theta\approx 1.1$) makes it challenging, if not impossible, to carry out 
a straight many-body calculation even in the already simplified lattice model. 
A further approximation may consist in focusing just on the 
four
mini bands, 
an approach
which requires 
to first identify their corresponding Wannier functions.  
Surprisingly, even
such a preliminary step turns out to be rather difficult and, to some extent, 
controversial.~\cite{Vafek,Koshino-Fu,Yuan-Fu,Senthil-1,Bernevig_topo,Vishwanath}
The scope of the present work is to shed light on this  debated issue. 

\section{Preliminary definitions and results}
\label{section II}
\begin{figure}
\centerline{\includegraphics[width=0.45\textwidth]{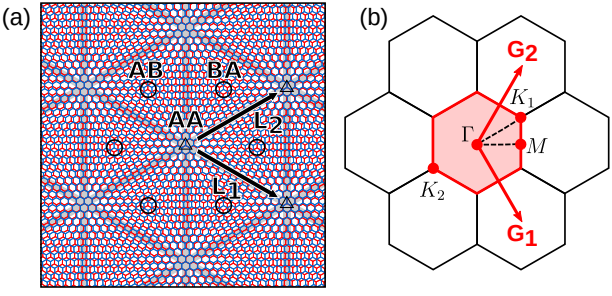}}
\caption{(a) Two graphene sheets rotated by a small angle (shown here for $\theta\approx 3.89^\circ$, while remaining calculations will be for $\theta = 1.08^\circ$) 
with respect to each other.
The emerging moir\'e pattern is highlighted by a grey shaded line 
and the predominant character of the stacking between the two layers, AA and AB (or BA), 
is indicated by black triangles and circles, respectively.
The triangular superlattice vectors $\textbf{L}_1$ and $\textbf{L}_2$ connects different AA zones.
(b) Mini Brillouin zone of tBLG. 
The high symmetry points $\Gamma,\bK_1,\bK_2,\bM$ are shown together 
with the reciprocal lattice vectors $\textbf{G}_1$ and $\textbf{G}_2$.}
\label{honeycomb-BZ}
\end{figure}

\begin{figure}[bht]
\centerline{\includegraphics[width=0.45\textwidth]{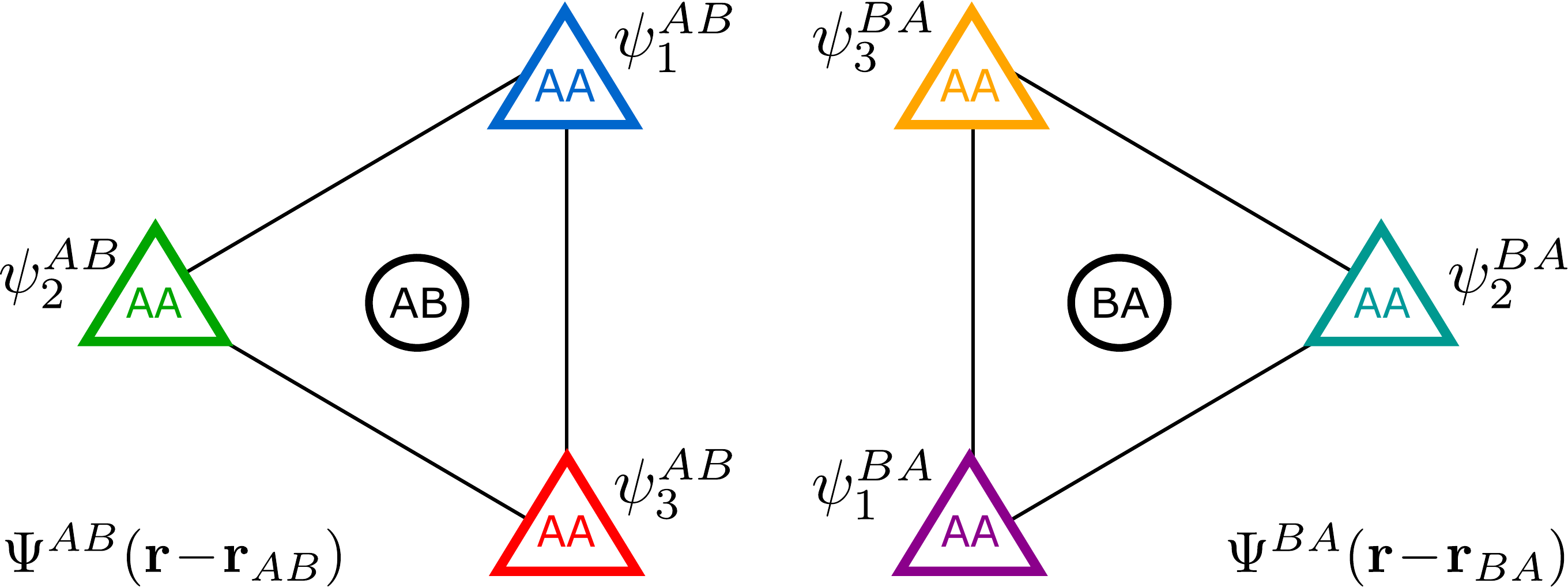}}
\caption{Pictorial view of the Wannier functions $\Psi^{AB}(\br-\br_{AB})$ and $\Psi^{BA}(\br-\br_{BA})$ centred at AB and BA sites, respectively. The triangles represent wavefunction components centred around 
the AA regions, while the combination of the three triangles defines the Wannier orbital, centred instead 
around AB, left, or BA, right. }
\label{Wannier}
\end{figure}

In Fig.~\ref{honeycomb-BZ}(a) we show two graphene layers rotated with
respect to the each other by a small angle. 
Due to the small misalignment between the graphene layers, a moir\'e 
pattern forms where regions characterized by local realizations of different stacking modes 
appear periodically within the bilayer.
Bernal-stacked regions (AB or BA) form an
honeycomb lattice  (black circles in Fig.~\ref{honeycomb-BZ}), while 
AA-stacked regions in the 
hexagon
centers 
form a triangular lattice 
(black 
triangles
in Fig.~\ref{honeycomb-BZ}).
If the twisted bilayer is obtained from AA stacking upon rotation around the center of two overlapping
basic graphene
hexagons, the point-group symmetry of the superlattice is $\text{D}_6$, which reduces to $\text{D}_3$
if, 
as we shall assume in the following, 
the rotation center is 
around a vertical C-C bond
~\cite{Senthil-2,Vafek} 
However, irrespective of the actual 
structural
symmetry group, there is wide consensus\cite{Bernevig_topo,Senthil-1,Koshino-Fu,Vafek} 
that a proper description of the band structure can be obtained 
by just assuming that the Wannier orbitals of the mini bands are centred 
on the AB and BA sites of the honeycomb moir\'e superlattice, 
even though their 
actual
weight  is 
mostly localized on the 
AA regions. 
For this reason we parametrize the Wannier orbitals 
$\Psi^{AB}(\br-\br_{AB})$ and $\Psi^{BA}(\br-\br_{BA})$
centred around the AB and BA sites with coordinates $\br_{AB}$ and 
$\br_{BA}$, respectively, through the 
functions $\psi^{AB}_i(\br-\bR_i)$ and $\psi^{BA}_i(\br-\bR'_i)$, $i=1,2,3$,
centred instead around the neighbouring AA sites with coordinates $\bR_i$ and $\bR'_i$ that are actually lattice sites of the triangular supercell, see Fig.~\ref{Wannier}.

In particular we shall assume that $\psi^{AB}_1$ and $\psi^{BA}_3$ are centred at the origin, 
taken to coincide with AA center $\bR_1=\bR'_3=\bnot$, so that $\bR_2=-\textbf{L}_1$, $\bR'_2=\textbf{L}_2$, $\bR_3=\bR'_1=\textbf{L}_1-\textbf{L}_2$, $\br_{AB}= (\textbf{L}_1-2\textbf{L}_2)/3$ and 
$\br_{BA}=(2\textbf{L}_1-\textbf{L}_2)/3$,
where $\textbf{L}_1$ and $\textbf{L}_2$ are the lattice vectors shown in Fig.~\ref{honeycomb-BZ}(a).\\
It follows that the most general Bloch function $\Phi_\bk(\br)$ 
can be written as 
\beal
&\Phi_\bk(\br) = \fract{1}{\sqrt{V}}\sum_\bR\!\Bigg(
u_\bk\,\esp{-i\bk\cdot(\bR+\br_{AB})}\,
\Psi^{AB}(\br-\br_{AB}-\bR) \\
&\qquad\qquad\quad  +  v_\bk\,\esp{-i\bk\cdot(\bR+\br_{BA})}\;\Psi^{BA}(\br-\br_{BA}-\bR)\Bigg)\\
&\quad= \fract{1}{\sqrt{V}}\sum_\bR
\esp{-i\bk\cdot\bR}\bigg(u_\bk\,\psi^{AB}_{\bk}(\br-\bR) 
+  v_\bk\,\psi^{BA}_{\bk}(\br-\bR)\bigg)\\
&\qquad \equiv \fract{1}{\sqrt{V}}\,\sum_\bR
\esp{-i\bk\cdot\bR}\;\phi_\bk(\br-\bR)
\,,
\label{Bloch}
\eal
where $\left|u_\bk\right|^2+\left|v_\bk\right|^2=1$, 
$V$ is the area,
and 
\bea
\psi^{AB}_{\bk}(\br) &=& \psi^{AB}_1(\br)\,\esp{-i\bk\cdot(\textbf{L}_1-2\textbf{L}_2)/3} \nonumber\\
&& \quad+ \psi^{AB}_2(\br)\,\esp{-i\bk\cdot(\textbf{L}_1+\textbf{L}_2)/3} \nonumber\\
&& \quad+ \psi^{AB}_3(\br)\,\esp{-i\bk\cdot(-2\textbf{L}_1+\textbf{L}_2)/3}\;,\label{AB}\\
\psi^{BA}_{\bk}(\br) &=&\psi^{BA}_1(\br)\,\esp{-i\bk\cdot(-\textbf{L}_1+2\textbf{L}_2)/3} 
\nonumber\\
&&\quad + \psi^{BA}_2(\br)\,\esp{-i\bk\cdot(-\textbf{L}_1-\textbf{L}_2)/3} 
\nonumber \\
&&\quad + \psi^{BA}_3(\br)\,
\esp{-i\bk\cdot(2\textbf{L}_1-\textbf{L}_2)/3}\;.\label{BA}
\eea
We note that, even though $\phi_\bk(\br-\bR)$ might be confused with the Wannier function centred in the triangular site $\bR$, yet it is not
so
 because of the explicit dependence upon momentum $\bk$.
In particular, under a symmetry transformation 
$\mathcal{G}$, such that $\br\to \br_G$ and $\bk\to\bk_G$, 
\beal
\mathcal{G}\Big(\Phi_\bk(\br)\Big) &= \fract{1}{\sqrt{V}}\,\sum_\bR\,
\esp{-i\bk_G\cdot\bR}\,\phi_\bk(\br_G-\bR) \,,
\eal
%
the outcome
simplifies only at the high-symmetry $\bk$-points, i.e., when $\bk_G\equiv \bk$ apart from a reciprocal lattice vector, in which case 
\be
\mathcal{G}\Big(\Phi_\bk(\br)\Big) = \Phi_\bk(\br_G)\,.\label{special-k}
\ee

In Fig.~\ref{honeycomb-BZ}(b) we show the first Brillouin zone, the reciprocal lattice vectors $\textbf{G}_1$ and 
$\textbf{G}_2$, as well as the high-symmetry $\bk$-points $\bGamma$, 
$\bK_1=(\textbf{G}_1+2\textbf{G}_2)/3$, $\bK_2=-\bK_1$ and $\bM=(\textbf{G}_1+\textbf{G}_2)/2$. 
The symmetry group G=D$_6$ is generated by C$_{3z}$, C$_{2z}$ and C$_{2x}$, while G=D$_3$ only 
by  C$_{3z}$ and C$_{2y}=\text{C}_{2z}\,\text{C}_{2x}$. 
The little group L at $\bGamma$
coincides with the full G, thus either D$_6$ or D$_3$,  while, at $\bK_1$ or $\bK_2$, L is generated 
only by C$_{3z}$ for both G=D$_6$ and G=D$_3$. 
It follows that the symmetry properties of the Bloch wavefunctions at $\bGamma$ 
can discriminate between 
G=D$_6$ and G=D$_3$, as we shall indeed show.  \\
Going back to the definitions \eqn{AB} and \eqn{BA}, we find for the high-symmetry points shown in 
Fig.~\ref{honeycomb-BZ}(b), 
\beal
\psi^{AB}_\bGamma(\br) &= \psi^{AB}_1(\br) +\psi^{AB}_2(\br) +\psi^{AB}_3(\br)\,,\\
\psi^{BA}_\bGamma(\br) &= \psi^{BA}_1(\br) +\psi^{BA}_2(\br) +\psi^{BA}_3(\br)\,,\label{Gamma}
\eal 
at $\bGamma$, while at $\bK_1$, 
\beal
\psi^{AB}_{\bK_1}(\br) &= \omega\,\bigg(\psi^{AB}_1(\br) +\omega\,\psi^{AB}_2(\br) +\omega^*\,\psi^{AB}_3(\br)\bigg)\,,\\
\psi^{BA}_{\bK_1}(\br) &= \omega^*\,\bigg(\psi^{BA}_1(\br) +\omega^*\,\psi^{BA}_2(\br) +\omega\,\psi^{BA}_3(\br)\bigg)\,,
\label{K}
\eal
and finally at $\bK_2$, 
\beal
\psi^{AB}_{\bK_2}(\br) &= \omega^*\,\bigg(\psi^{AB}_1(\br) +\omega^*\,\psi^{AB}_2(\br) +\omega\,\psi^{AB}_3(\br)\bigg)\,,\\
\psi^{BA}_{\bK_2}(\br) &= \omega\,\bigg(\psi^{BA}_1(\br) +\omega\,\psi^{BA}_2(\br) +\omega^*\,\psi^{BA}_3(\br)\bigg)\,,
\label{-K}
\eal
where $\omega=\esp{i 2\pi/3}$.  \\
For later convenience, we recall how the different symmetry operations act in tBLG. We 
 write the coordinate of a carbon atom as $\br=(x,y,z)\equiv(\br_{||},z)$, where $z=-1$ indicates the bottom layer \#1 while $z=+1$ the upper one \#2. The planar coordinate $\br_{||}$ 
may belong to sublattice A or B of each graphene layer, as well as to the AB or BA 
sublattice regions of the superlattice. It follows that C$_{3z}$ changes neither $z$ nor the sublattice index, 
both of the original lattice, A or B, as well as of the superlattice, AB or BA. 
On the contrary, under C$_{2z}$,  $z\leftrightarrow z$, 
$A\leftrightarrow B$ and $AB\leftrightarrow BA$. Finally, under C$_{2x}$, 
$z\leftrightarrow -z$, $A\leftrightarrow B$, while AB and BA are invariant. 
%

\section{Lattice relaxation and tight binding calculation of the \MakeLowercase{t}BLG bandstructure}
\label{sec:MD}

\subsection{Model and simulation protocol}
\label{subsec:MDprotocol}

The above symmetry analysis strictly holds only for an idealized tBLG obtained by a rigid rotation of the layers without atomic relaxation.  However, there 
is strong evidence of
a substantial lattice relaxation, especially at small twist angles,\cite{Yoo_180403806,Zhang2018} which  needs to be accounted for to get physically reliable results.  \\

We thus performed lattice relaxations via classical molecular dynamics simulations 
using state-of-the-art force-fields. 
We select  a few angles in the range of $\theta \approx 1^{\circ}$-$1.5^{\circ}$, 
at which perfectly periodic (commensurate) structures can be built.~\cite{Dos_SantosPRL} 
We consider an aligned bilayer ($\theta=0^{\circ}$) in the AA stacking configuration, and 
rotate the upper layer around a carbon atom, 
which corresponds to a
type II structure~\cite{Senthil-2} with only $\text{D}_3$ symmetry.
The carbon-carbon intralayer interactions are modelled 
via the second generation REBO potential.\cite{Brenner-JPhysCondMat2002}
The interlayer interactions are instead modelled 
via the Kolmogorov-Crespi (KC) potential,\cite{Kolmogorov-PPRB2005} 
using the recent parametrization of Ref.~\onlinecite{Ouyang-NanoLett2018}. 
The starting intralayer carbon-carbon distance is set equal to $a_0=1.3978$ \AA\,, 
corresponding to the equilibrium bond length of the adopted REBO potential, 
giving a lattice parameter of $a\approx2.42$ \AA. 
Geometric optimizations are performed using the FIRE algorithm.\cite{Bitzek-PRL2006}
The atomic positions are relaxed toward equilibrium until total force acting on 
each atom, $F_i=|-\nabla_{{\bf r}_i}(V^{\rm KC}_{\rm inter}+V^{\rm REBO}_{\rm intra})|$, 
become less than $10^{-6}$ eV/atom.
It is important to stress
that during the relaxation the system is not constrained 
to preserve any particular symmetry. 
 \begin{table}
 \renewcommand{\arraystretch}{1.3}
 \begin{tabular}{  c  c  c }
 \hline
 \hline
          & interlayer dist. & $\Delta{\mathrm \varepsilon}$\\
          & (\AA) & (meV/atom)\\
 \hline
  AB &  3.39 & 0 \\
 \hline
  SP &  3.42 & 0.74\\
 \hline
  AA &  3.61 & 4.70\\
 \hline
 \hline
 \end{tabular}
 \caption{The equilibrium interlayer distance and the corresponding total energy 
  of aligned 
  ($\theta=0^{\circ}$) graphene bilayers at various stacking modes, specified in the first column. 
  Energies are measured relative to that of the optimal AB stacking. 
Results obtained by initialling shifting the relative (x,y) centers-of mass of the two layers, and then relaxing.
For the case of AB stacking, a full relaxation of the bilayer was performed. For the case of AA or SP 
  stacking, only the $z$ coordinate of all atoms was relaxed, while the in plane ($x$,$y$) coordinates
  were held fixed. This prevented the bilayer from falling into the AB global minimum, thus preserving the
  initial stacking.
 }
  \label{Table-1}

\end{table}
\begin{figure*}[htb]
\centerline{\includegraphics[width=1.0\textwidth]{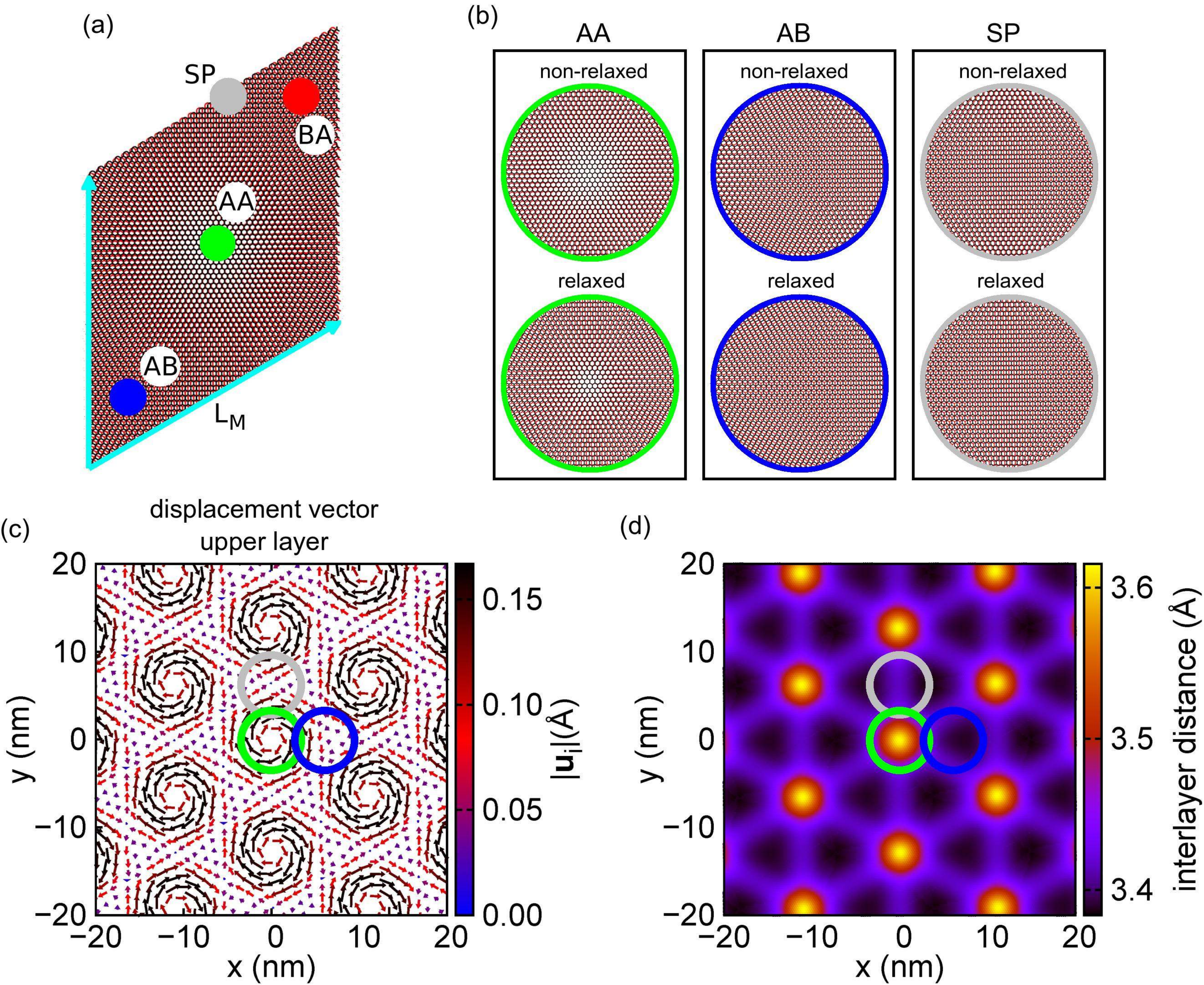}}
\caption{\label{fig:MD}
          (a) The supercell of a tBLG at $\theta\approx1.08^{\circ}$ used in simulations, obtained upon rotating 
          a bilayer initially in the AA stacking configuration
around a vertical C-C bond ($\text{D}_3$ structure). Arrows show the 
primitive lattice vectors, of length $L_{M}$, of the triangular moir\'e superstructure. 
          Green, grey, red and blue circles mark the regions of AA, SP, AB and BA stacking, respectively. 
          (b) Local structure 
          before and after relaxation around the center of the AA, SP and AB regions. 
          (c) Displacement field showing the in-plane deformations of the upper layer. 
          The displacement vectors \{${\bf u}_i$\} go from the equilibrium position 
          of the carbon atoms in the non-relaxed configuration to the corresponding position 
          in the fully relaxed structure. 
          Only few vectors are shown for clarity, magnified by a factor of ten. 
          (d) Colored map showing the local interlayer distance. 
          The colored circles reported in panels (c) and (d) 
          correspond to the samples of panel (b).}
\end{figure*}

\subsection{Results: optimized geometry of magic angle tBLG}
\label{subsec:MDresults}
Fig.~\ref{fig:MD}(a) shows the supercell of tBLG at $\theta\approx1.08^{\circ}$, before
relaxation, corresponding to a triangular superlattice of period\cite{Sboychakov}
$L_{M}=|L_1|=|L_2|=\frac{a}{2\sin(\theta/2)}\approx13$ nm and, as mentioned before, D$_3$ symmetry. 
Examining
different directions, areas of energetically least favourable AA stacking, see Table~\ref{Table-1},  gradually 
turn into energetically more favourable saddle point (SP) regions 
or most favourable AB and BA stacking regions.
As previously reported,\cite{Yoo_180403806,Uchida_PRB,Zhang2018,Dai_Nanolett,Nam_Koshino_PRB,Juricic,Yazyev-1} 
after full relaxation the AA regions shrink while the area of the Bernal-stacked 
regions expand (see Fig.~\ref{fig:MD}(b)). 
This is achieved via small in-plane deformations characterized by a displacement field 
that rotates around the center of the AA domains (see Fig.~\ref{fig:MD}(c)), 
respectively counterclockwise and clockwise in the upper and lower layer. 
We note that such distortions lead to negligible local lattice compressions/expansions, 
corresponding to variations $<0.03$\% of the stiff carbon-carbon bond length
relative to the equilibrium value. On the other hand, the large difference between the 
equilibrium interlayer distances of the AA and AB stacking (see Table~\ref{Table-1}) leads to 
significant out-of-plane buckling deformations, genuine "corrugations"
of the graphene layers, that form protruding bubbles in correspondence of the AA regions. This is 
clearly shown in Fig.~\ref{fig:MD}(d),  where the colour map of the local interlayer distance, 
shows an overall increase of $\sim0.2$ \AA\, from Bernal  AB (blue circle) to the AA region (green circle).
We end by emphasising that the relaxed structure does not exhibit any manifest point-group symmetry, despite 
its initial D$_3$ symmetry
before relaxation.
Na\"{i}vely, one should then conclude that all the symmetry analysis of the previous section is 
unjustified and 
meaningless. 
We shall show below that this is not the case.

\subsection{Tight-binding electronic structure calculations}

While the above discussion focused on a specific supercell at $\theta\approx1.08^{\circ}$, 
qualitatively similar results were obtained for other angles, too.
We emphasize that 
out-of-plane deformations, 
significant at small magic angles,  
have important effects on the electronic structure of the system. 
Indeed, as can be seen from Fig.~\ref{bande}(b), 
where the tight-binding band structure is calculated for the fully relaxed structure, 
the flat bands are now well separated from the rest by an $\approx 45-50$ meV gap, 
consistent with experiment \cite{Herrero-1,Yankowitz,Herrero-2}, and larger 
than the gap obtained allowing only in-plane displacements.\cite{Nam_Koshino_PRB}

Tight-binding calculation details are standard. 
Denoting the position within the unit cell of atom $i$ as $\br_i$ we can write the tight-binding Hamiltonian as:
\beal
\hat{\mathcal{H}}=\sum_{i,j} \Big(t\big(\br_i-\br_j\big)\ket{i}\bra{j} + \ \text{H.c.}\Big)\,,\label{Hamiltonian}
\eal
where $t(\br_i-\br_j)$ is the hopping amplitude which is computed using the Slater-Koster formalism:~\cite{SK_integral}
\beal
t(\mathbf{d})= V_{pp\sigma}(d)\bigg[\frac{\textbf{d}\cdot \textbf{e}_z}{d}\bigg]^2 \!\!\!+V_{pp\pi}(d)
\bigg[1-\Big(\frac{\textbf{d}\cdot \textbf{e}_z}{d}\Big)^2\bigg]\,,
\eal
where $\mathbf{d}= \br_i-\br_j$, $d=|\mathbf{d}|$, and $\textbf{e}_z$ is the unit vector in the direction perpendicular to the graphene planes.
The out-of-plane ($\sigma$)  and in-plane ($\pi$) transfer integrals are:
\beal
V_{pp\sigma}(x)=V_{pp\sigma}^0 \esp{-\frac{x-d_0}{r_0}}\;\;\;\;V_{pp\pi}(x)
=V_{pp\pi}^0 \esp{-\frac{x-a_0}{r_0}}
\eal
 where $V_{pp\sigma}^0=0.48\;eV$ and $V_{pp\pi}^0=-2.7\; eV$ are values chosen to reproduce ab-initio dispersion curves in AA and AB stacked bilayer graphene, $d_0=3.344 \text{\AA}$ is the starting inter-layer distance, 
$a_0=1.3978 \text{\AA}$ is the intralayer carbon-carbon
distance, as previously defined, and $r_0=0.184~a$ 
is the decay length, in units of the lattice parameter.\cite{Trambly,Nam_Koshino_PRB}
Although the hopping amplitude decreases exponentially with distance, we
found
that upon 
setting
even a fairly large cutoff $r_c$, important features of the band structure are spoiled. 
An example is the degeneracy at the $K_{1(2)}$ points, 
which we find to be fourfold, up to our numerical accuracy, keeping all hopping amplitudes that are nonzero within machine precision, while it is fully lifted using a cutoff as large as $r_c\approx 4a_0$. 

In addition, we 
assumed 
the carbon $\pi$-orbitals 
to be
oriented along $\textbf{e}_z$, 
while in reality they are oriented along the direction locally perpendicular to the relaxed graphene sheet, 
no longer
flat. However, since the out of plane distortions varies smoothly along the moir\'e pattern, we 
checked that the misorientation of the orbitals with respect to the $z$ axis 
are lower than $\approx 0.1-0.01^\circ$, and have no noticeable effect on the band structure.

\section{Symmetry analysis of the Bloch functions}
\begin{figure}[htb]
\centerline{\includegraphics[width=0.475\textwidth]{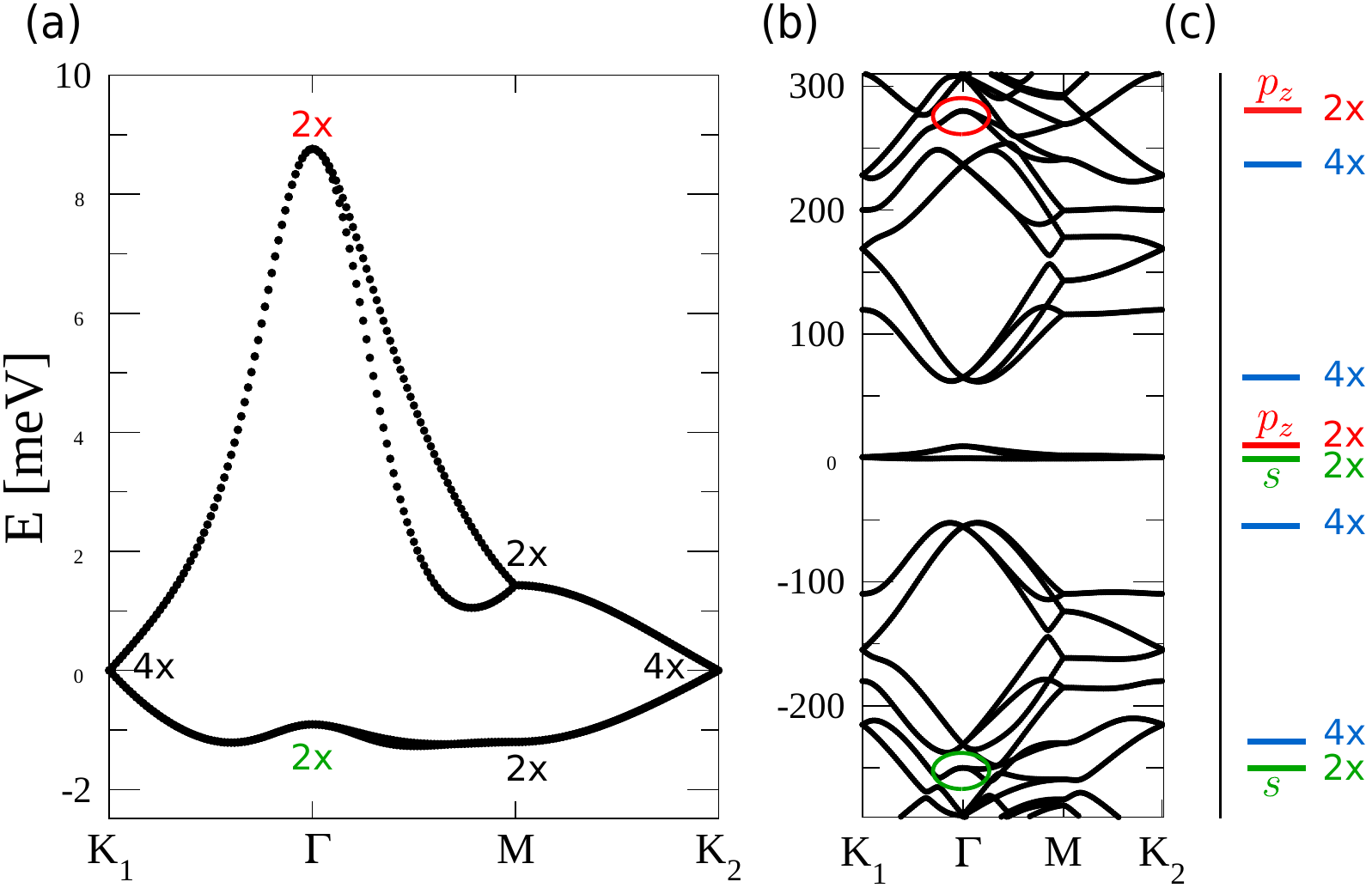}
}
\caption{Band structure at twist angle $1.08^\circ$ of the relaxed tBLG. (a) A zoom-in of the band 
structure showing only the four mini bands, where labels indicate their degeneracy      
at the high symmetry points. (b) The full band structure. 
The two circles indicate the $s$ (below) and $p_z$ (up) doublets 
used to construct the Wannier orbitals. (c) Level spectrum and degeneracy at the $\bGamma$ point. The label $s$ and $p_z$ refer to the symmetry under C$_{2x}$, see the text.
}
\label{bande}
\end{figure}
In Fig.~\ref{bande} we show the band structure around the neutrality point. 
In Fig.~\ref{bande}(a), we plot just the four 
mini bands, which are well separated from the others, see Fig.~\ref{bande}(b). 
We also indicate the degeneracy 
at the high symmetry points. In particular, at $\bK_1$ and $\bK_2$ 
we find that all four bands are degenerate within our numerical accuracy, 
while they are split into two doublets at $\bGamma$ and $\bM$. 
In Fig.~\ref{bande}(c), we show the level spectrum at the $\bGamma$ point, including the degeneracy of each level.\\ 

\begin{figure}[bht]
\centerline{\includegraphics[width=0.475\textwidth]{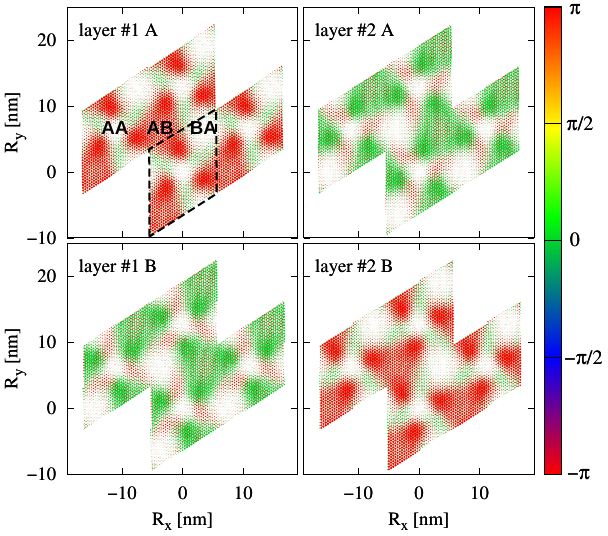}
}
\caption{Layer ($\#1$-$\#2$) and sublattice (A-B) components of one state within the lowest-energy doublet at $\bGamma$ in the flat bands. The colour of each point indicates its complex phase, while its size is a measure of its square modulus. Each unit cell (black dashed line in top left panel) has been replicated 3 times to improve visibility.  This eigenstate is invariant under $\text{C}_{3z}$, even with respect to $\text{C}_{2x}$ and odd under $\text{C}_{2z}$.}
\label{Gamma-1}
\end{figure}

Even though the relaxed lattice has no manifest point symmetry, we shall 
still assume,
arbitrarily for now,
either D$_3$ or D$_6$ symmetry 
retaining
the formalism of 
Sec.~\ref{section II}. The comparison with the tight-binding 
results will decide upon the validity of that assumption. \\ 
Since the Wannier functions are centred at the vertices of the hexagons, 
where the symmetry is C$_3$ irrespective of the 
global symmetry being D$_6$ or D$_3$, 
one could 
be tempted to
rationalize\cite{Vafek,Koshino-Fu} the 
miniband $\Gamma$ point double
degeneracy 
as due to
two different $\Psi^{AB}$, as well as $\Psi^{BA}$, see Fig.~\ref{Wannier}, which transform as the 
two-dimensional irreducible representation
of C$_3$. We find that this assumption is not correct in our case. In Fig.~\ref{Gamma-1} we show the wavefunction of one of the two states 
within the lower doublet at $\bGamma$. 
It is 
visually evident, and also
confirmed numerically, that a 
mini-band Bloch wavefunction 
at $\Gamma$ 
is instead invariant under C$_{3z}$, which implies that the Wannier functions 
must transform as one of the singlet irreps of C$_3$. 
The same is true for all the other three Bloch functions 
which we do not show. Assuming therefore that all the 
Wannier functions are invariant under C$_{3z}$, we can 
parametrize the functions $\psi^{AB}_i(\br)$, $i=1,2,3$, of 
Fig.~\ref{Wannier} as follows
\beal
\psi^{AB}_1(\br) &= A(\br) + E_{+1}(\br) + E_{-1}(\br)\,,\\
\psi^{AB}_2(\br) &= A(\br) + \omega\,E_{+1}(\br) + \omega^*\,E_{-1}(\br)\,,\\
\psi^{AB}_3(\br) &= A(\br) + \omega^*\,E_{+1}(\br) + \omega\,E_{-1}(\br)\,,\label{new-wannier}
\eal 
where $A(\br)$ is invariant under C$_3$, while $E_{\pm 1}(\br)$ transforms 
with eigenvalue $\omega^{\pm1}=\esp{\pm i 2\pi/3}$. 
Recalling that $\psi^{AB}_{n+1}(\br-\textbf{L}_2)=\text{C}_3(\psi^{AB}_n(\br-\textbf{0}))$ 
($n=1,2,3$ and $n+3=n$), one can readily show that 
the Wannier function $\Psi^{AB}(\br)$ shown in Fig.~\ref{Wannier} 
is indeed invariant under C$_{3z}$. 
Similarly, 
for $\psi^{BA}_i(\br)$ we introduce the functions $A'(\br)$ and $E_{\pm 1}'(\br)$. 
It follows that the Eqs.~\eqn{Gamma}~and~\eqn{K} simplify 
to    
\beal
\psi^{AB}_\bGamma(\br) &= 3A(\br)\,,\\
\psi^{AB}_{\bK_1}(\br) &= 3\omega\,E_{-1}(\br)\,,\\
\psi^{AB}_{\bK_2}(\br) &= 3\omega^*\,E_{+1}(\br)\,,
\label{new-bloch-AB}
\eal
for AB, and 
\beal
\psi^{BA}_\bGamma(\br) &= 3A'(\br)\,,\\
\psi^{BA}_{\bK_1}(\br) &= 3\omega^*\,E'_{+1}(\br)\,,\\
\psi^{BA}_{\bK_2}(\br) &= 3\omega\,E'_{-1}(\br)\,,
\label{new-bloch-BA}
\eal
for BA. 
Therefore, studying the Bloch functions at the different high-symmetry points gives direct access to $A(\br)$ as well as $E_{\pm 1}(\br)$, as we show in what follows. 
 
\subsection{Bloch functions at $\bGamma$} 
We start our analysis from the $\bGamma$ point. 
Looking again at Fig.~\ref{Gamma-1}, one 
notes that
the Bloch functions have negligible amplitude in the AA zones, 
being mostly localized in AB/BA,\cite{Mellado} 
and thus the Wannier orbitals cannot be localized in AA only. 
Most importantly, 
one finds that the Bloch function is not only
invariant under C$_{3z}$, but also 
possesses well defined symmetry properties under C$_{2z}$ and C$_{2x}$, 
specifically it is odd under the former, 
cf. panel layer \#1 A with panel layer \#1 B, 
and even under the latter, 
cf. panel \#1 A with panel layer \#2 B. 
Similarly, the other state within the lower doublet is still even under C$_{2x}$, 
but also even under C$_{2z}$. 
That 
doublet thus transforms with respect to C$_{2x}$ as an $s$-orbital. 
On the contrary, the upper doublet is odd under C$_{2x}$, thus transforming as a $p_z$-orbital, 
one state being even and the other odd under C$_{2z}$. We thus conclude that close to the charge neutrality point the effective symmetry group is 
actually D$_6$,\cite{Senthil-2,Bernevig_topo,Vishwanath}
and hence contains also C$_{2z}$, 
even if
the relaxed structure lacks any point symmetry. 

We stress 
in addition
that the double degeneracy of the mini-bands at $\bGamma$ is generically 
not to be expected even assuming D$_6$ symmetry. 
The accidental degeneracy is 
due to the fact that the coupling between the Dirac points, which originally belonged to different layers and correspond to the same momentum $\bK_1$ or $\bK_2$ in the reduced Brillouin zone, effectively vanishes
at small twist angles,\cite{MacDonald-PNAS2011} 
even though symmetry does not prohibit this 
coupling to be finite. This phenomenon corresponds to an additional emergent symmetry, 
dynamical in nature (some textbooks would call it accidental), 
often referred as valley charge conservation $U_v(1)$ symmetry.   \cite{Senthil-1,Bernevig_topo}.  
\begin{figure}[tb]
\centerline{\includegraphics[width=0.475\textwidth]{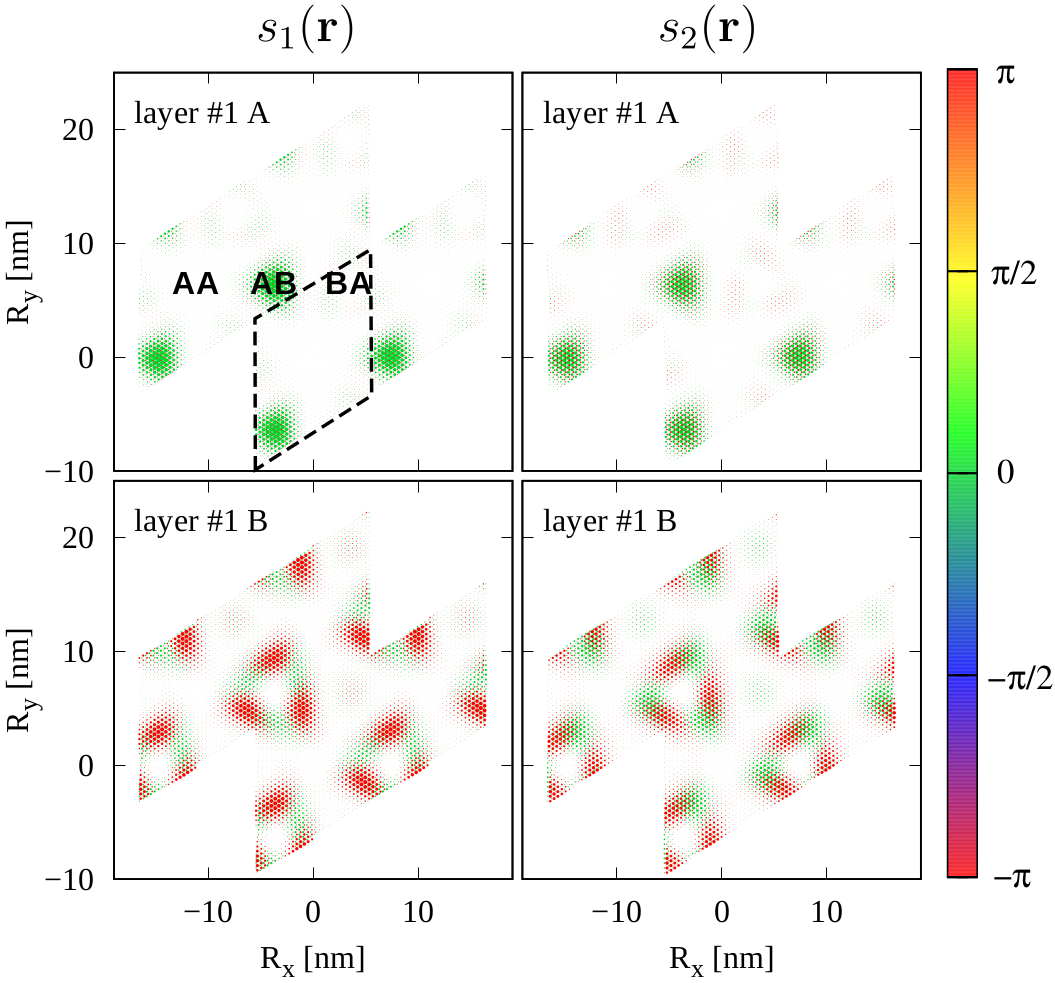}}
\caption{Layer $\#1$ and sublattice (A-B) components of $s_1(\br)$ (left panel) and $s_2(\br)$ (right panel). The colour of each point indicates its complex phase, while its size is a measure of its square modulus. Each unit cell (black dashed line in top left panel) has been replicated 3 times to improve visibility.}
\label{S12}
\end{figure}

\begin{figure}[b]
\centerline{\includegraphics[width=0.475\textwidth]{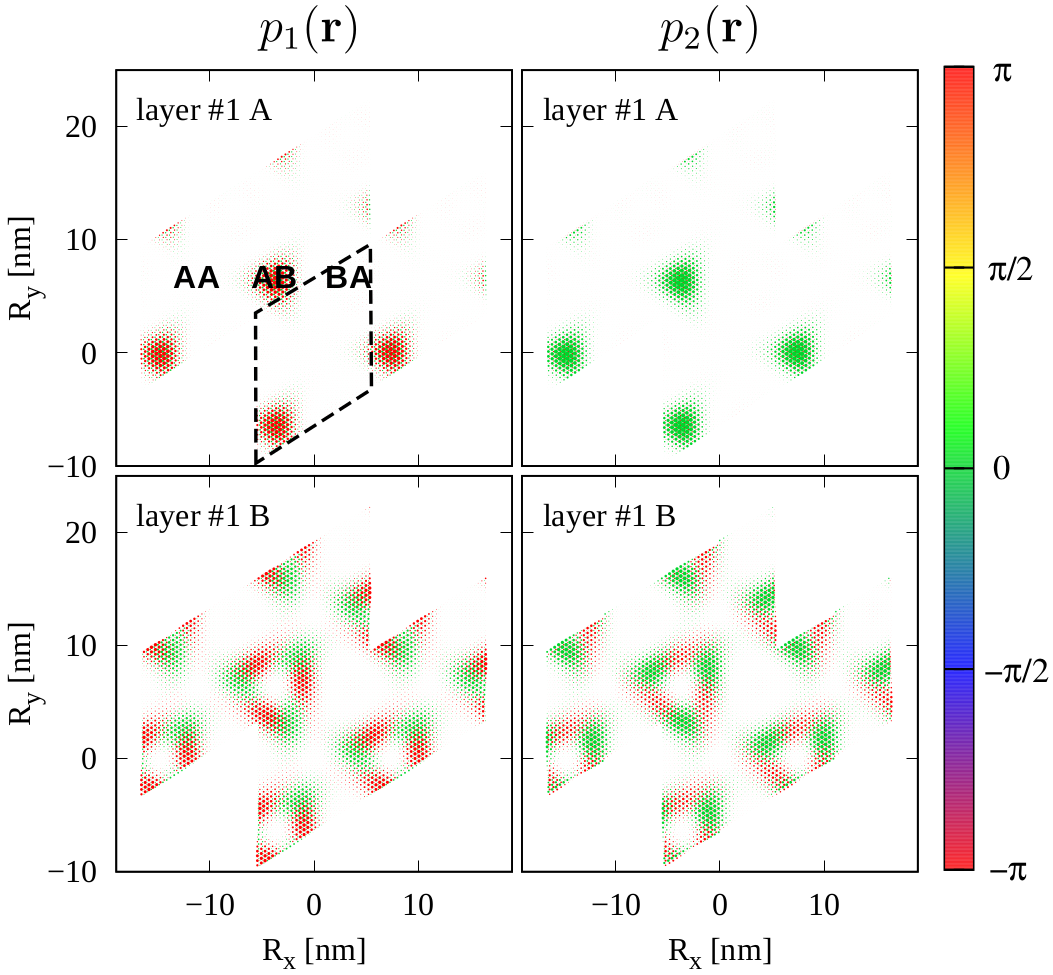}}
\caption{Layer $\#1$ and sublattice (A-B) components of $p_1(\br)$ (left panel) and $p_2(\br)$ (right panel). The colour of each point indicates its complex phase, while its size is a measure of its square modulus. Each unit cell (black dashed line in top left panel) has been replicated 3 times to improve visibility }
\label{P12}
\end{figure}

If so, AB and BA being equivalent, the function $\phi_\bk(\br)$, 
see Eqs.~\eqn{Bloch},~\eqn{Gamma},~\eqn{new-bloch-AB}~and~\eqn{new-bloch-BA}, 
at $\bGamma$ can be written as 
\be
\phi_\bGamma(\br) = 3A(\br)\pm 3A'(\br)\,,\label{intermediate}
\ee
i.e., sum or difference of the AB and BA components. 
Since the two combinations cannot be degenerate, 
in order to describe the band structure we need at least two different $s$-like 
and two different $p_z$-like orbitals for each sublattice AB or BA. 
It thus follows that there must be two additional doublets above or below the flat-bands, 
one of $s$-type and another of $p_z$-type, both invariant under C$_{3z}$. 
As can be seen in Fig.~\ref{bande}(b) and (c), 
above the flat-bands at $\bGamma$ there are two fourfold degenerate levels 
that actually transform as the 
two-dimensional irreducible representation, 
and hence are not invariant under $\text{C}_{3z}$. 
The next two states (upper red circle) have instead the right symmetry properties, 
i.e., they are invariant under three-fold rotations and have well defined parity, 
actually odd, under $\text{C}_{2x}$ 
(one being even and one odd with respect to $\text{C}_{2z}$). 
This doublet is therefore the partner of the $p_z$-doublet in the mini band. 
The same holds in the lower energy bands (lower green circle). 
With the only difference that the doublet is 
now even under $\text{C}_{2x}$, 
hence it is the partner of the $s$-doublet in the mini band. 
Let us focus for instance on the two $s$-orbitals, 
and denote $3A(\br)$ either as $s_1(\br)$ or $s_2(\br)$, 
and similarly $3A'(\br)$ as $s'_1(\br)$ or $s'_2(\br)$.  
We assume that the $s$-doublet below the mini bands corresponds 
to the AB+BA combination, hence, through Eqs.~\eqn{new-bloch-AB}~and~\eqn{new-bloch-BA}, 
\beal
\phi^{(1+)}_\bGamma(\br) &= s_1(\br) + s_1'(\br)\,,\\
\phi^{(2+)}_\bGamma(\br) &= s_2(\br) + s_2'(\br)\,.
\eal
If $\phi^{(1+)}_\bGamma$ is chosen to be even under C$_{2z}$, 
so that $\phi^{(2+)}_\bGamma$ is odd, then 
\beal
s_1' &= \mathcal{C}_{2z}\big(s_1\big)\,,\qquad 
s_2' = -\mathcal{C}_{2z}\big(s_2\big)\,.
\eal
The $s$-doublet within the mini bands must therefore be the AB-BA combination
\beal
\phi^{(1-)}_\bGamma(\br) &= s_1(\br) - s_1'(\br)\,,\\
\phi^{(2-)}_\bGamma(\br) &= s_2(\br) - s_2'(\br)\,,
\eal
so that $\phi^{(1-)}_\bGamma$ is odd under C$_{2z}$, while 
$\phi^{(2-)}_\bGamma$ even. It follows that taking either the sum or the difference between two states belonging to different $s$-doublets with opposite parity under C$_{2z}$, we should find wavefunctions centred either in AB or BA. This is indeed the case. In Fig.~\ref{S12} we show the layer \#1 sublattice components of $s_1(\br)$, left panel, and $s_2(\br)$, right panel. The components on layer \#2 can be obtained through C$_{2x}$, and the functions $s_1'(\br)$ and $s_2'(\br)$ on the sublattice BA 
through C$_{2z}$. 
We can repeat a similar analysis to find the two $p_z$-type 
functions, $p_1(\br)$ and $p_2(\br)$, which are shown in 
Fig.~\ref{P12}. 

We conclude 
by stressing
that the same symmetry partners 
of the mini band levels at $\bGamma$ are 
no less than
$300~\text{meV}$ 
away
from them, and in between there are several states with different symmetry. However, as soon as we move away from $\bGamma$ all those states will be coupled to each other by the Hamiltonian, and thus a description in terms only of few of them is hardly possible. 

\subsection{Bloch functions at $\bK$}

\begin{figure}
\centerline{\includegraphics[width=0.45\textwidth]{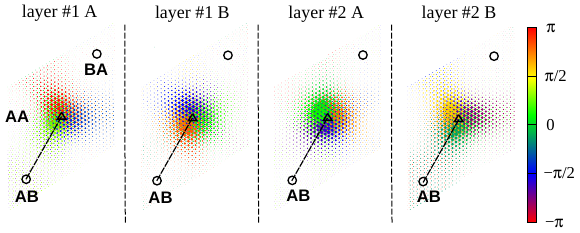}}
\caption{Layer and sublattice components in the unit cell of one of the two degenerate Bloch functions at $\bK_1$ whose Wannier orbitals are centred on AB.}
\label{KAB-1}
\end{figure}

At the high-symmetry points $\bK_1$ and $\bK_2=-\bK_1$ the AB and BA Wannier functions are effectively decoupled and degenerate. However, the outcome of numerical diagonalization is a generic linear combination of the degenerate levels. Therefore, in order to identify AB and BA components,  we introduced a small perturbation in the Hamiltonian that makes AB and BA inequivalent while preserving the 
$\text{C}_{3z}$ symmetry:
\begin{equation}
V(\textbf{r})=-\sum_{j=1}^3 2 V_0 \text{sin}(\textbf{g}_j \cdot \textbf{r}),
\end{equation}
where
$\textbf{g}_1=\textbf{G}_1$, 
$\textbf{g}_2=\textbf{G}_2$, 
$\textbf{g}_3=-\textbf{G}_1-\textbf{G}_2$, and $V_0\approx~1~\;\mu$eV.
This function is
maximum in AB, minimum in BA and  
zero in AA. By doing so, the fourfold degenerate states at $\bK_{1/2}$ are split by a tiny gap (less than $0.2~\;\mu$eV) in two doublets, the lower/upper one composed by Bloch states that are combination of BA/AB Wannier orbitals only.
In such a way, we can directly obtain the 
proper lattice-symmetric 
functions $E_{\pm 1}(\br)$ and $E'_{\pm 1}(\br)$ through Eqs.~\eqn{new-bloch-AB} and \eqn{new-bloch-BA}.  
Since there are four states at $\bK_{1/2}$, there will be two different $E_{+1}(\br)$,and similarly
for all the other components. In Fig.~\ref{KAB-1} we show the layer and sublattice components 
of one of  the two degenerate Bloch functions at $\bK_1$ centred on AB. We note that this Bloch functions transforms under C$_{3z}$ as the expected $E_{-1}(\br)$, see Eq.~\eqn{new-bloch-AB}. We did check 
that all other Bloch functions at $\bK_1$ and $\bK_2$ are compatible 
with Eqs.~\eqn{new-bloch-AB}~and~\eqn{new-bloch-BA}. 

\section{Conclusions}
We presented
a theoretical and numerical analysis of 
the electronic structure associated with the fully relaxed geometric structure of
a twisted bilayer graphene at small twist angles,  which 
must be relevant for the intriguing behavior 
observed in recent experiments.~\cite{Herrero-1,Herrero-2,Yankowitz}
In particular, with state-of-the-art techniques, we model both the in-plane and out-of-plane 
atomic relaxations, and we show that they play a crucial role 
in reproducing the experimentally observed 
one-electron band gaps. 
By performing an extensive study of the Bloch eigenfunctions at the high symmetry points, 
we are able to single out the symmetry properties 
and in fact the rather subtle nature
of the corresponding Wannier orbitals. 
The results are consistent with a 
$\text{D}_6$ symmetry, which emerges despite the absence of an {\it a priori} lattice structure
point group symmetry, as well as with a valley charge-conservation $U_v(1)$. %
These emerging symmetries are robust features of small angle twisted bilayer graphene. 
Moreover, even though the flat bands are well separated from the rest, 
in order to simultaneously describe the physics 
at both the $\bK$ and $\bGamma$ points, 
one necessarily has to consider an enlarged set of Wannier orbitals, 
at least eight but most likely much more. 
The impact of these results in our understanding of the observed phenomena in twisted graphene bilayers will be the subject of a future work. \\

\section*{Acknowledgments}

We thank P.~Jarillo-Herrero for useful discussions. 
D.~M. acknowledges the fellowship from the Sackler Center for Computational Molecular and Materials Science at Tel Aviv University, and from Tel Aviv University Center for Nanoscience and Nanotechnology.
A.~V. acknowledges financial support from the Austrian Science Fund (FWF) 
through the Erwin Schr\"odinger fellowship J3890-N36.
A.~A., A.~V., and M.~C. also acknowledge financial support from MIUR PRIN 2015 (Prot. 2015C5SEJJ001) and SISSA/CNR project ``Superconductivity, Ferroelectricity and Magnetism in bad metals" (Prot. 232/2015).
M.~F. acknowledges support by the European Union under H2020 Framework Programs, ERC Advanced Grant No. 692670 ``FIRSTORM''. 
E.~T. acknowledges support by the European Union under FP7  ERC Advanced Grant No.320796   "MODPHYSFRICT". \\

\textit{Note added:} After completion of the present study, we
became aware of a recent preprint,\cite{Woo_arxiv} which also reports 
a relaxed structure with some similar features of the model shown here.

\bibliographystyle{apsrev}

\end{document}